\documentstyle[12pt]{article}
\newcommand{\address}[1]%
{\centerline{\small\it #1}}
\renewcommand{\title}[1]{\begin{center}%
{\large\bf #1}\end{center}\par\bigskip}
\renewcommand{\author}[1]{\centerline{#1}}
\renewcommand{\maketitle}{}
\newcommand{\pacs}[1]{}
\newcommand{\narrowtext}{}

\newcommand{\gtrsim}
{\mathrel{\raisebox{-2.8pt}{\mbox%
{$\stackrel{\textstyle >}{\sim}$}}}}
\begin{document}
\title{The second law
of Thermodynamics
as a theorem in quantum 
mechanics}
\author{Hal Tasaki}
\address{Department of Physics, 
Gakushuin University,
Mejiro, Toshima-ku, Tokyo 171, JAPAN}
\maketitle
%%%%%%%%%%%%%%%%%%%%%%%%%%%%%%%%%%%%%%%%%%%%
\begin{abstract}
We treat a quantum mechanical system with
certain general properties which are expected 
to be common in macroscopic
quantum systems.
Starting from a {\em pure}\/ initial state
(which may not describe an equilibrium)
in which energy is mildly concentrated 
at a single value,
we consider a time evolution determined 
by a time-dependent Hamiltonian as a 
model of an adiabatic operation in
thermodynamics.
We take a family of operations with 
the same procedure and various
``waiting times.''
Then the {\em minimum work principle}
is rigorously proved
for almost all choices of the waiting time.
\end{abstract}
%%%%%%%%%%%%%%%%%%%%%%%%%%%%%%%%%%%%%%%%%%%
%%%%%%%%%%%%%%%%%%%%%%%%%%%%%%%%%%%%%%%%%%%
%%%%%%%%%%%%%%%%%%%%%%%%%%%%%%%%%%%%%%%%%%%
\narrowtext
To develop a microscopic
understanding of the second law of thermodynamics 
\cite{LY},
which is one of the most perfect macroscopic
laws in physics, has been an
unsolved fundamental problem.
If one puts aside the problem of
``thermalization'' and starts from equilibrium
states in the sense of statistical mechanics,
then one can reasonably derive the second 
law in some situations \cite{der1,der2}.

In the present paper, we make a further step 
by discussing a 
rigorous derivation of the second law
directly from quantum mechanics
without referring to statistical mechanics.
More precisely, we model 
{\em adiabatic operation}\/ in thermodynamics
as a time evolution according to a 
time dependent Hamiltonian \cite{op}
in a closed quantum mechanical system.
We assume that the system has certain general 
properties (such as the existence of
extensive entropy, nondegeracy and non-resonance
of the energy eigenvalues)
which we expect to be common in macroscopic
systems.
We take the initial state to be a {\em pure}\/
state in which the energy is concentrated
at a single value, but not too sharply.
We consider a family of operations with the same
procedure but with various ``waiting times.''
Then we prove, for large enough systems,
an inequality corresponding to the {\em minimum
work principle}\/ for almost all choices of the
waiting time.
The initial state may or may not correspond
to a macroscopic equilibrium state.
Thus we are dealing also with the problem
of ``thermalization'' (although in an indirect
and incomplete manner).

\paragraph*{Setup and main results:}
We consider a quantum mechanical system 
characterized by a single parameter
\( V>0 \) (the volume) which can be made as
large as one wishes.
For each \( V \) we fix two Hamiltonians 
\( H_{\rm init} \) and
\( H_{\rm fin} \) which have (infinite)
discrete eigenvalues
\( E_{j} \) and \( E'_{j} \) 
(\( j=1,2,\ldots \)), 
respectively.
We assume \( E_{j}<E_{j+1} \) and 
\( 0\le E'_{j}\le E'_{j+1} \)
for any \( j \).
(For simplicity, we do not make explicit
the \( V \)-dependence of the Hamiltonians,
the energy eigenvalues, and the eigenstates.)
We have assumed the energy eigenvalues of 
\( H_{\rm init} \) to
be nondegenerate.
We further assume that these 
eigenvalues satisfy the
{\em non-resonance condition}\/, 
i.e.,
\( E_{j}-E_{k}=E_{\ell}-E_{m}\ne0 \)
implies \( j=\ell \) and \( k=m \).
We expect that a generic 
Hamiltonian of a macroscopic
quantum systems satisfy 
these conditions \cite{cond}.

We assume that 
\( H_{\rm init} \)
and \( H_{\rm fin} \) have well-behaved
thermodynamic limits as 
\( V\to\infty \) in the following sense.
Define the number of states 
\( \Omega_{V}(E) \) to be
the number of eigenvalues 
\( E_{j} \) of \( H_{\rm init} \)
such that \( E_{j}\le E \).
We assume that there are 
\( V \)-independent constants 
\( C_{1}>0 \),
\( C_{2}>0 \), \( a \), \( b \), and 
a smooth increasing function 
\( s(\cdot) \),
and the number of states satisfy
\begin{equation}
	C_{1}V^{a}
	\exp\left[V\,s\left(
	\frac{E}{V}\right)\right]
	\le
	\Omega_{V}(E)
	\le
	C_{2}V^{b}
	\exp\left[V\,s\left(
	\frac{E}{V}\right)\right],
	\label{eq:OEe}
\end{equation}
which is an expected behavior
in a macroscopic system with 
the (infinite volume) entropy
\( s(\epsilon)=
\lim_{V\to\infty}V^{-1}
\log\Omega_{V}(V\epsilon) \).
We also assume that there are
\( V \)-independent smooth functions
\( f(\cdot) \), \( g_{1}(\cdot) \) and
\begin{equation}
	\left|\frac{E'_{j}}{V}-
	f\left(\frac{E_{j}}{V}\right)
	\right|
	\le
	\frac{1}{V}\,
	g_{1}\left(\frac{E_{j}}{V}\right),
	\label{eq:EVfEV}
\end{equation}
for any \( j \).
(Throughout the present paper,
\( g_{i}(\cdot) \) is a function which appears
as an unimportant coefficient of a 
``small'' term.)

In order to describe an adiabatic operation of
an outside agent to the system,
we choose (for a fixed \( V \)) an arbitrary
time-dependent Hamiltonian \( H_{0}(\cdot) \)
such that \( H_{0}(0)=H_{\rm init} \)
and  \( H_{0}(T)=H_{\rm fin} \).
For an arbitrary \( \tau\ge0 \), 
consider the same operation
executed after a ``waiting time'' of \( \tau \),
which is described by the Hamiltonian 
\( H_{\tau}(\cdot) \) defined
as 
\begin{equation}
	H_{\tau}(t)=
	\cases{
	H_{\rm init}
	&
	for \( 0\le t\le\tau \);
	\cr
	H_{0}(t-\tau)
	&
	for \( \tau\le t\le T+\tau \).
	}
	\label{eq:Htau}
\end{equation}
It is essential in our approach to investigate 
the
family of operations with a fixed 
\( H_{0}(\cdot) \)
and various ``waiting time'' \( \tau \).

For a fixed \( \tau\ge0 \),
let \( U_{\tau}(\cdot) \) be the solution of 
\( i\partial U_{\tau}(t)/\partial t =
H_{\tau}(t)U_{\tau}(t)\),
and define \( U_{\tau}=U_{\tau}(T+\tau) \)
which is the unitary operator representing the 
whole operation.

Let \( \varphi_{j} \) and 
\( \varphi'_{j} \)
be the normalized eigenstates of 
the Hamiltonians
\( H_{\rm init} \) and \( H_{\rm fin} \), 
respectively,
with eigenvalues \( E_{j} \) and 
\( E'_{j} \).
Then we define a unitary operator 
\( U_{\rm slow} \)
by 
\( U_{\rm slow}\varphi_{j}=\varphi'_{j} \).
Recalling the ``adiabatic theorem'' 
in quantum mechanics,
one may interpret \( U_{\rm slow} \) 
as describing the
time evolution in an operation 
where the Hamiltonian
changes from \( H_{\rm init} \) to
\( H_{\rm fin} \) in an infinitely slow manner.

We assume that the system is initially 
in a normalized
pure state 
\( \varphi_{\rm init} \), and 
require that, 
in \( \varphi_{\rm init} \),
{\em the energy
is concentrated around a single 
value, but not too sharply}\/.
(It is trivial to extend all the results to 
mixed initial states.)
More precisely we expand 
it as 
\( \varphi_{\rm init}
=\sum_{j}\xi_{j}\varphi_{j} \),
where \( \xi_{j} \) are complex coefficients,
and assume that 
\begin{equation}
	\xi_{j}=0
	\quad
	\mbox{if}
	\quad
	\left|
	\frac{E_{j}}{V}-\epsilon_{0}
	\right|\ge
	C_{3}V^{-\delta}
	\label{eq:xi0}
\end{equation}
where \( \epsilon_{0} \), 
\( C_{3}>0 \), and
\( 0<\delta<1 \) are (\( V \)-independent)
constants,
and
\begin{equation}
	|\xi_{j}|^2
	\le
	\frac{1}{\Omega_{V}(\epsilon_{0}V)}
	\label{eq:xiupper}
\end{equation}
for any \( j \).
Note that this upper bound for 
\( |\xi_{j}|^2 \) is quite mild
since the number of allowed basis states
(which is 
\( \Omega_{V}(\epsilon_{0}V
+C_{3}V^{1-\delta})
-\Omega_{V}(\epsilon_{0}V
-C_{3}V^{1-\delta})\))
is much larger than 
\( \Omega_{V}(\epsilon_{0}V) \)
if \( V \) is large.
(From (\ref{eq:OEe}), we find it is 
larger by a factor of
\( \exp(C_{3}s'(\epsilon_{0})V^{1-\delta}) \).)
There is a huge freedom in the choice of
\( \xi_{j} \) within the restrictions
(\ref{eq:xi0}) and (\ref{eq:xiupper}).
In particular, we are {\em not}\/ 
assuming anything
like all the eigenstates with energies in a 
finite range appear
with equal weights.
We expect that generic 
macroscopic states
(which, unlike Schr\"{o}dinger's cat states,
have more or less determined energies) 
automatically posses similar 
properties \cite{artificial}.

Now suppose that the 
system is initially in the 
state \( \varphi_{\rm init} \), and
one measures the energy (described by 
\( H_{\rm fin} \))
after the operation.
The energy expectation value after the
``slow'' operation is 
\begin{equation}
	E_{\rm slow}=
	\langle\varphi_{\rm init},
	U_{\rm slow}^{-1}H_{\rm fin}U_{\rm slow}
	\varphi_{\rm init}\rangle
	=\sum_{j}|\xi_{j}|^2E'_{j}.
	\label{eq:Eslow}
\end{equation}
and that after the
operation described by \( H_{\tau} \) is
\begin{equation}
	E_{\tau}=
	\langle\varphi_{\rm init},
	U_{\tau}^{-1}H_{\rm fin}U_{\tau}
	\varphi_{\rm init}\rangle.
	\label{eq:Etau}
\end{equation}
Then our result is as follows.

\bigskip\noindent
{\bf Theorem:}
Take an arbitrary family of models
with  \( H_{\rm init} \), \( H_{\rm fin} \),
\( H_{0}(\cdot) \), and \( \varphi_{\rm init} \)
satisfying all the conditions stated above.
Then for sufficiently large \( V \),
there exist \( \tau_{\rm max} \)
and a subset \( I\subset[0,\tau_{\rm max}] \)
such that
\begin{equation}
	\frac{|I|}{\tau_{\rm max}}
	\ge
	1-\exp\left[
	-\frac{V\,s(\epsilon_{0})}{2}\right],
	\label{eq:muI}
\end{equation}
where \( |I| \) denotes the
``total length''
of the set \( I \),
and for any \( \tau\in I \) we have
\begin{equation}
	\frac{E_{\tau}}{V}\ge
	\frac{E_{\rm slow}}{V}
	-g_{2}(\epsilon_{0})V^{-\delta}
	\label{eq:main}
\end{equation}
where \( g_{2}(\cdot) \) is a 
(\( V \)-independent)
function which depend only on 
\( C_{1} \), \( C_{2} \), \( a \), \( b \),
\( \delta \), \( s(\cdot) \), \( f(\cdot) \),
and \( g_{1}(\cdot) \).

\bigskip
The theorem says that the inequality
(\ref{eq:main}) holds unless one happens to 
choose the
``waiting time'' \( \tau \) 
from a very exceptional
set 
\( [0,\tau_{\rm max}]\backslash I \).
Since the proportion of the exceptional set
is exponentially small in \( V \), 
there is practically no chance of observing the
violation of (\ref{eq:main}) when 
\( V \) is large
(provided that one takes the waiting
time long enough).

Now the inequality (\ref{eq:main}) implies
\begin{equation}
	\frac{W_{\tau}}{V}\ge
	\frac{W_{\rm slow}}{V}
	-g_{2}(\epsilon_{0})V^{-\delta},
	\label{eq:MWP}
\end{equation}
where 
\( W_{\tau}=
E_{\tau}-
\langle\varphi_{\rm init},H_{\rm init}
\varphi_{\rm init}\rangle \)
and
\( W_{\rm slow}=
E_{\rm slow}-
\langle\varphi_{\rm init},H_{\rm init}
\varphi_{\rm init}\rangle \)
are the works done by the agent to the system
during the operations.
When \( V \) is large, the inequality
(\ref{eq:MWP}) becomes
\( W_{\tau}/V\gtrsim W_{\rm slow}/V \),
which is nothing but the
{\em minimum work principle}\/
\cite{W}
(for a closed system) in thermodynamics
\cite{next}.
Since the minimum work principle is
expected to hold for an arbitrary 
operation applied on a thermodynamic
equilibrium state,
the present result shows 
(although in an indirect manner)
that an equilibrium is attained
after sufficiently long
``waiting time'' \cite{note:tau}.

We note that the minimum work principle for
closed systems is one of the fundamental
forms of the second law for 
simple systems, and
other forms (such as the Kelvin's principle
or the law of entropy increase)
can be derived from it by suitable
thermodynamic arguments.
Moreover this form of the second law
does not rely on definitions of
heat or entropy, which are always delicate.
It depends solely on the notion of energy transfer
which is the ultimate object to be studied in
thermodynamics.

\paragraph*{Proof:}
From  (\ref{eq:Eslow}), (\ref{eq:xi0}), 
and (\ref{eq:EVfEV}), we see
\begin{equation}
	\frac{E_{\rm slow}}{V}
	\le
	f(\epsilon_{0}+C_{3}V^{-\delta})
	+\frac{g_{1}(\epsilon_{0}
	+C_{3}V^{-\delta})}
	{V},
	\label{eq:VfE}
\end{equation}
which implies
\begin{equation}
	\frac{E_{\rm slow}}{V}
	\le
	\frac{\bar{E}}{V}
	+g_{3}(\epsilon_{0})V^{-\delta},
	\label{eq:EEg}
\end{equation}
where \( \bar{E}=Vf(\epsilon_{0}) \),
and \( g_{3}(\cdot) \) is a 
\( V \)-independent function.

Since \( H_{\rm fin} \) is unbounded in general, 
we introduce
a bounded operator
\( \tilde{H}=\bar{E}(1-P)+H_{\rm fin}P \)
where \( P \) is the orthogonal projection
onto the space spanned by 
\( \varphi'_{j} \)
with \( j \) such that 
\( E'_{j}\le\bar{E}=Vf(\epsilon_{0}) \).
\( \tilde{H} \) behaves exactly as 
\( H_{\rm fin} \) for
energies lower than \( \bar{E} \),
and behaves as a constant otherwise.
Let 
\( \tilde{E}_{\tau}
=\langle\varphi_{\rm init},
U_{\tau}^{-1}\tilde{H}U_{\tau}
\varphi_{\rm init}\rangle\).
Since \( H_{\rm fin}\ge\tilde{H} \),
we have 
\( E_{\tau}\ge\tilde{E}_{\tau} \).
We shall prove (\ref{eq:main}) with
\( E_{\tau} \) replaced by 
\( \tilde{E}_{\tau} \),
which automatically leads to the desired result
for \( E_{\tau} \).

By noting that 
\( U_{\tau}=U_{0}
\exp(-iH_{\rm init}\tau) \),
we see that
\begin{eqnarray}
	\tilde{E}_{\tau}
	&=&
	\sum_{j,j'}\xi_{j}^*\xi_{j'}
	\langle\varphi_{j},
	U_{\tau}^{-1}\tilde{H}
	U_{\tau}\varphi_{j'}\rangle
	\nonumber\\
	&=&
	\sum_{j,j'}\xi_{j}^*\xi_{j'}
	\exp[i(E_{j}-E_{j'})\tau]
	\langle\varphi_{j},
	U_{0}^{-1}\tilde{H}
	U_{0}\varphi_{j'}\rangle.
	\label{eq:Etil}
\end{eqnarray}
For a function \( f_{\tau} \) 
of \( \tau \),
we denote its ``long-(waiting-)time'' average as
\begin{equation}
	\overline{f_{\tau}}
	=
	\lim_{S\to\infty}\frac{1}{S}
	\int_{0}^Sd\tau\,f_{\tau}.
	\label{eq:overlinef}
\end{equation}
Since the assumption of nondegeneracy implies
\( \overline{e^{i(E_{j}-E_{j'})\tau}}
=\delta_{j,j'} \),
we have
\begin{eqnarray}
	\overline{\tilde{E}_{\tau}}
	&=&
	\sum_{j}|\xi_{j}|^2
	\langle\varphi_{j},
	U_{0}^{-1}\tilde{H}
	U_{0}\varphi_{j}\rangle
	\nonumber\\
	&=&
	\sum_{j,k}
	|\xi_{j}|^2
	\langle\varphi_{j},
	U_{0}^{-1}\varphi'_{k}\rangle
	\langle\varphi'_{k},
	\tilde{H}\varphi'_{k}\rangle
	\langle\varphi'_{k},
	U_{0}\varphi_{j}\rangle
	\nonumber\\
	&=&
	\sum_{j,k}
	|\xi_{j}|^2\alpha_{j,k}\tilde{E}'_{k},
	\label{eq:Etil2}
\end{eqnarray}
where \( \alpha_{j,k}=
|\langle\varphi_{j},U_{0}^{-1}
\varphi'_{k}\rangle|^2\),
and
\( \tilde{E}'_{k}=
\langle\varphi'_{k},\tilde{H}
\varphi'_{k}\rangle
=\min\{E'_{k},\bar{E}\} \).
Because of the unitarity, 
\( (\alpha_{j,k}) \) is a doubly stochastic
matrix \cite{Bhatia}, i.e., it satisfies 
\( 0\le\alpha_{j,k}\le1 \) and
\( \sum_{j}\alpha_{j,k}
=\sum_{k}\alpha_{j,k}=1 \).
Let \( j(\cdot) \)  be a permutation of 
integers such that 
\( |\xi_{j(\ell)}|\ge
|\xi_{j(\ell+1)}| \).
Then one can easily prove \cite{ineq}
that 
\begin{equation}
	\overline{\tilde{E}_{\tau}}
	=
	\sum_{j,k}
	|\xi_{j}|^2\alpha_{j,k}\tilde{E}'_{k}
	\ge
	\sum_{k=1}^\infty
	|\xi_{j(k)}|^2\tilde{E}'_{k}.
	\label{eq:Etil3}
\end{equation}
Let 
\( \Omega_{-}=
\Omega_{V}(V\epsilon_{0}
-C_{3}V^{1-\delta}) \).
Then, by using (\ref{eq:Etil3}),
(\ref{eq:EVfEV}),
\( \bar{E}=Vf(\epsilon_{0}) \),
(\ref{eq:xiupper}), (\ref{eq:OEe}),
and (\ref{eq:EEg}),
we can estimate the average 
\( \overline{\tilde{E}_{\tau}} \) as
\begin{eqnarray}
	\overline{\tilde{E}_{\tau}}
	&\ge&
	\sum_{k=\Omega_{-}}^\infty
	|\xi_{j(k)}|^2\tilde{E}'_{k}
	\nonumber\\
	&\ge&
	\tilde{E}'_{\Omega_{-}}
	\sum_{k=\Omega_{-}}^\infty
	|\xi_{j(k)}|^2
	\nonumber\\
	&\ge&
	\{
	V\,f(\epsilon_{0}-C_{3}V^{-\delta})
	-g_{1}(\epsilon_{0}-C_{3}V^{-\delta})
	\}
	\sum_{k=\Omega_{-}}^\infty
	|\xi_{j(k)}|^2
	\nonumber\\
	&\ge&
	\{\bar{E}-g_{4}(\epsilon_{0})
	V^{1-\delta}\}
	\left(
	1-
	\sum_{k=1}^{\Omega_{-}-1}|\xi_{j(k)}|^2
	\right)
	\nonumber\\
	&\ge&
	\{\bar{E}-g_{4}(\epsilon_{0})
	V^{1-\delta}\}
	\left\{
	1-\frac{\Omega_{V}(V\epsilon_{0}
	-C_{3}V^{1-\delta})}
	{\Omega_{V}(V\epsilon_{0})}
	\right\}
	\nonumber\\
	&\ge&
	\{\bar{E}-g_{4}(\epsilon_{0})
	V^{1-\delta}\}
	\left\{
	1-\frac{C_{2}}{C_{1}}V^{b-a}
	\exp[
	-V\{s(\epsilon_{0})
	-s(\epsilon_{0}-C_{3}V^{-\delta})\}
	]
	\right\}
	\nonumber\\
	&\ge&
	\{\bar{E}-g_{4}(\epsilon_{0})
	V^{1-\delta}\}
	\left\{
	1-\frac{C_{2}}{C_{1}}V^{b-a}
	\exp[
	-g_{5}(\epsilon)V^{1-\delta}
	]
	\right\}
	\nonumber\\
	&\ge&
	\bar{E}-g_{6}(\epsilon_{0})
	V^{1-\delta}
	\nonumber\\
	&\ge&
	E_{\rm slow}-g_{7}(\epsilon_{0})
	V^{1-\delta},
	\label{eq:Etil4}
\end{eqnarray}
for sufficiently large \( V \),
where \( g_{i}(\cdot) \) are 
\( V \)-independent
functions.

In order to convert the above bound for the
average into 
information about 
\( \tilde{E}_{\tau} \) 
itself \cite{old}, we evaluate the variance
and follow the standard argument in the proof
of Chebyshev inequality \cite{Feller}.

From (\ref{eq:Etil}),
we see that
\begin{equation}
	(\tilde{E}_{\tau})^2
	=
	\sum_{j,k,\ell,m}
	\xi_{j}^*\,\xi_{k}\,
	\xi_{\ell}^*\,\xi_{m}\,
	e^{i(E_{j}-E_{k}
	+E_{\ell}-E_{m})\tau}\,
	h(j,k)\,h(\ell,m),
	\label{eq:Esq}
\end{equation}
where 
\( h(j,j')=
\langle\varphi_{j},U_{0}^{-1}\tilde{H}
U_{0}\varphi_{j'}\rangle \).
From the non-resonance condition for 
the spectrum of
\( H_{\rm init} \),
we see that the average
\( \overline{e^{i(E_{j}-E_{k}
+E_{\ell}-E_{m})\tau}} \)
is equal to 1 when \( j=k \), 
\( \ell=m \)
or \( j=m \), \( k=\ell \),
and is vanishing otherwise.
This means that
\begin{equation}
	\overline{(\tilde{E}_{\tau})^2}
	=\left\{
	\sum_{j}|\xi_{j}|^2h(j,j)
	\right\}^2
	+
	\sum_{j\ne j'}
	|\xi_{j}|^2|\xi_{j'}|^2h(j,j')h(j',j).
	\label{eq:Esq15}
\end{equation}
Noting that the first term in the
right-hand side is equal to
\( \left(
	\overline{\tilde{E}_{\tau}}
	\right)^2 \), we find that
\begin{eqnarray}
	\overline{
	\left(\tilde{E}_{\tau}-
	\overline{\tilde{E}_{\tau}}\right)^2
	}
	&=&
	\overline{(\tilde{E}_{\tau})^2}
	-
	\left(
	\overline{\tilde{E}_{\tau}}
	\right)^2
	\nonumber\\
	&=&
	\sum_{j\ne j'}
	|\xi_{j}|^2|\xi_{j'}|^2h(j,j')h(j',j)
	\nonumber\\
	&\le&
	\left\{\max_{j'}|\xi_{j'}|^2\right\}
	\sum_{j,j'}
	|\xi_{j}|^2
	\langle\varphi_{j},U_{0}^{-1}\tilde{H}
	U_{0}\varphi_{j'}\rangle
	\langle\varphi_{j'},U_{0}^{-1}\tilde{H}
	U_{0}\varphi_{j}\rangle
	\nonumber\\
	&=&
	\left\{\max_{j'}|\xi_{j'}|^2\right\}
	\sum_{j}
	|\xi_{j}|^2
	\langle\varphi_{j},U_{0}^{-1}
	\tilde{H}^2
	U_{0}\varphi_{j}\rangle
	\nonumber\\
	&\le&
	\frac{\bar{E}^2}
	{\Omega_{V}(\epsilon_{0}V)},
	\label{eq:Esq2}
\end{eqnarray}
where we used (\ref{eq:xiupper})
and \( \tilde{H}\le\bar{E} \).
Since 
\( \left(\tilde{E}_{\tau}-
\overline{\tilde{E}_{\tau}}\right)^2 \)
is continuous in \( \tau \),
we see that for sufficiently large 
\( \tau_{\rm max} \)
\begin{equation}
	\frac{1}{\tau_{\rm max}}
	\int_{0}^{\tau_{\rm max}}
	d\tau
	\left(\tilde{E}_{\tau}-
	\overline{\tilde{E}_{\tau}}\right)^2
	\le
	\frac{2\,\bar{E}^2}
	{\Omega_{V}(\epsilon_{0}V)}.
	\label{eq:varupper}
\end{equation}
Define
\begin{equation}
	I=\{\tau\,\,|\,\,
	0\le\tau\le\tau_{\rm max},
	|\tilde{E}_{\tau}
	-\overline{\tilde{E}_{\tau}}|
	\le
	C_{4}V^{1-\delta}\}.
	\label{eq:Idef}
\end{equation}
Let \( \chi[\mbox{true}]=1 \) and
\( \chi[\mbox{false}]=0 \).
Then, by noting that
\( \chi[|x|\ge x_{0}]\le(x/x_{0})^2 \),
and using (\ref{eq:varupper}) 
and (\ref{eq:OEe}),
we see that
\begin{eqnarray}
	1-\frac{\mu[I]}{\tau_{\rm max}}
	&=&
	\frac{1}{\tau_{\rm max}}
	\int_{0}^{\tau_{\rm max}}
	d\tau
	\,
	\chi[|\tilde{E}_{\tau}-
	\overline{\tilde{E}_{\tau}}|\ge
	C_{4}V^{1-\delta}]
	\nonumber\\
	&\le&
	\frac{1}{\tau_{\rm max}}
	\int_{0}^{\tau_{\rm max}}
	d\tau
	\left\{
	\frac{\tilde{E}_{\tau}-
	\overline{\tilde{E}_{\tau}}}
	{C_{4}V^{1-\delta}}
	\right\}^2
	\nonumber\\
	&\le&
	\left(
	\frac{\bar{E}}{V}
	\right)^2
	\frac{2\,V^{2\delta}}
	{(C_{4})^2\,\Omega_{V}(\epsilon_{0}V)}
	\nonumber\\
	&\le&
	\{f(\epsilon_{0})\}^2
	\frac{2\,V^{2\delta-a}}
	{C_{1}(C_{4})^2}
	\exp[-V\,s(\epsilon_{0})]
	\nonumber\\
	&\le&
	\exp\left[
	-\frac{V\,s(\epsilon_{0})}{2}\right],
	\label{eq:final}
\end{eqnarray}
where the final bound is valid for
sufficiently large \( V \).
This is the bound (\ref{eq:muI}) 
for the size of the
exceptional set.
Recalling (\ref{eq:Etil4}),
we have proved the
desired theorem with
\( g_{2}(\cdot)=g_{7}(\cdot)+C_{4} \).

%%%%%%%%%%%%%%%%%%%%%%%%%%%%%%%%%%%%%%%%%%%
%%%%%%%%%%%%%%%%%%%%%%%%%%%%%%%%%%%%%%%%%%%
\bigskip
It is a pleasure to thank
Elliott Lieb,
Takayuki Miyadera,
Hiroshi Nagaoka,
Shin-ichi Sasa,
Akira Shimizu,
and
Jakob Yngvason
for 
useful discussions on related topics.
%%%%%%%%%%%%%%%%%%%%%%%%%%%%%%%%%%%%%%%%%%%
%%%%%%%%%%%%%%%%%%%%%%%%%%%%%%%%%%%%%%%%%%%


\begin{thebibliography}{10}
\bibitem{op}
We imagine that an outside agent changes some
of the controllable parameters
(such as external fields or the positions of
walls) in the Hamiltonian
according to a fixed procedure.
We note that a general adiabatic operation in
thermodynamics \cite{LY} may 
not simply fit into 
such a scheme.
In reality an agent is even allowed to make
an observation and change the procedure
according to information thus obtained.
\bibitem{cond}
Obviously many simple noninteracting 
systems do not
satisfy these conditions.
We nevertheless expect that 
the degeneracies and the 
resonances in such
models are removed as soon as 
nontrivial interactions are turned on.
We also note that it is possible to
allow some amount of degeneracies
and resonances by making appropriate
modifications in the proof.
\bibitem{artificial}
We note that a rather artificial nature of
our conditions is not really 
necessary for our proof.
With more careful estimates,
we can treat models with much 
milder conditions.
\bibitem{next}
We expect, for not too radical operations,
\( U_{\tau}\varphi_{\rm init} \) possesses properties
similar to that we imposed on 
\( \varphi_{\rm init} \),
and hence our derivation of the minimum work
principle can be extended to the next 
operation where 
\( U_{\tau}\varphi_{\rm init} \) 
plays the role of initial state.
(But it seems that proofs of this claim should be done 
for specific models.)
\bibitem{W}
In order to give a thermodynamic interpretation,
we are implicitly assuming that the works are
extensive and have small fluctuation.
The inequality itself is of course valid
without the assumption.
\bibitem{note:tau}
This suggests that the role of
the waiting time is 
to let the system relax to an equilibrium.
We note, however, that the introduction of
\( \tau \) is necessary even when the
initial state \( \varphi_{\rm init} \) 
already corresponds to an equilibrium
(from a macroscopic point of view).
This is because, for any give
 \( \varphi_{\rm init} \),
one can construct a special
(probably highly unphysical and hard-to-realize)
unitary transformation which violates the
second law.
By considering operations with various \( \tau \),
we are excluding such a transformation.
\end{thebibliography}
\end{document}